\documentstyle[twoside,fleqn,espcrc2]{article}
\input{epsf}
\newcommand{\half}{\frac{1}{2}}
\newcommand{\threehalfs}{\frac{3}{2}}
\newcommand{\bc}{\begin{center}}
\newcommand{\ec}{\end{center}}

\newcommand{\ga}{\gamma}
\newcommand{\de}{\delta}

\newcommand{\be}{\[}
\newcommand{\ee}{\]}

\newcommand{\bea}{\begin{eqnarray}}
\newcommand{\eea}{\end{eqnarray}}
\newcommand{\ba}{\begin{array}}
\newcommand{\ea}{\end{array}}
\newcommand{\bn}{\begin{enumerate}}
\newcommand{\en}{\end{enumerate}}

\newcommand{\ssm}{supersymmetric standard model}

\newcommand{\pa}{\partial}

\newcommand{\npb}{{Nucl.\ Phys.\ }{\bf B}}

\newcommand{\spak}{\hspace*{.4in}}
\newcommand{\vr}{\vspace*{-0.14in}}
\newcommand{\vrl}{\vspace*{-0.07in}}
\title{The Mass Spectrum of a Static Adjoint Particle}
\author{M Foster and C Michael\address{      Division of Theoretical Physics,
                         Department of Mathematical Sciences,\\
                         University of Liverpool,	
                         PO Box 147, Liverpool L69 3BX, UK.},
{\em UKQCD Collaboration}
}
\begin{document}
\begin{abstract}

The bound states of fermions in the adjoint representation are of
interest in supersymmetric models. We investigate the energy spectrum of
the simplest -- the gluino-gluon bound states -- on several lattices in
the quenched approximation. We use a static approximation for the gluino
propagator. We find continuum limits of the splitting between the few
lowest states, with the energy difference between the two lowest states
of $354 \pm 9$ MeV.
 \vr
\end{abstract}

\maketitle

\section*{Motivation}  

The \ssm \space enriches the conventional particle spectrum in all
sectors. There is motivation to study states in strongly coupled N=1
SQCD, where the particle mass spectrum must be determined by
non-perturbative techniques. As well their relevance  for
supersymmetric phenomenology, these bound states, in R-parity conserving
theories, are of cosmological interest as dark matter
candidates. In addition to this, recent detector searches for such
particles have sought to eradicate the window for theories with a light
gaugino sector  \cite{farrah}.

In a low temperature regime we do not expect gluino loops to contribute
significantly to the propagation of a state. We study states containing
gluinos using quenched lattices and a static gluino propagator
approximation. The systematic error from quenching the quarks will be 
dominant. One of the simplest  of these states is the  bound states of
gluons and a gluino.  They can be constructed by measuring the
propagation of a single heavy gluino coupled to the Yang Mills vacuum.  
  
\section*{The Static Approximation}

The contribution to the QCD action of a heavy fermion is given by
 \be 
S=\int d^4x \bar{\psi}i\ga^{\mu}(\pa_{\mu}+igA_{\mu})\psi -M_Q \bar{\psi}\psi.
 \ee 
 The propagator for such fields is the Greens function of the
corresponding wave equation. In the limit $M_Q \longrightarrow \infty$,
we can discard space-like components of the covariant derivative and
note that the group contribution is given by the phase evolution of the
fermion as it moves. This provides us with a propagator for a static
fermion;
 \be 
K=\de^3(x^i-x^{\prime i}) \Pi(U^4) P_{\pm}e^{\mp M_Q(x_4-x^{\prime}_4)},
 \ee
 where P is the helicity projection operator and the sign depends on the
direction of particle evolution.

 We see that this can be extended to encompass fermions in the adjoint
representation simply by demanding the replacement of the link variables
$U^4_{ij}$ with elements of SU(3) in the adjoint representation; the
real $8 \times 8$ $G^4_{ab}$ matrices. These we construct by observing
that the combination of fundamental elements with the generators
Tr$(U\lambda^aU^{\dagger}\lambda^b)$ satisfy all the group requirements.

\section*{The Cubic Group}

We need to create a gluon field coupled to the static gluino propagator
in order to measure the energy from the correlation $C(t)$ with
Euclidean time:
 \be <C>=\Sigma_n<X^{\dagger}(t)|n><n|X(0)>e^{-E_nt}. \ee
 Here $X$ is the creation operator for the state, and $n$ is a
complete set of intermediate propagating states. 

For maximal signal and minimal contamination from other states we choose
$X$ to be irreducible representations of the cubic group $O\bigotimes
Z_2\bigotimes Z_2$\cite{var} such that $\mid  <\! X|n \!> \mid^2$ is both
large and orthogonal to other states.  

In the continuum limit states are labelled by $J^{PC}$. In a lattice
calculation the spatial rotational symmetry is broken down to that of
$O$ and we recover the spin content of $O$ by subduction. The Spin
content of $O$ is given in Table~\ref{spins} .

 \begin{table}[ht]
\vrl
\caption{Irreducible representations of $O$}
\label{spins}
\begin{tabular}{ccc}
 \hline
  Representation &   Dimension & Spin Content \\ 
 \hline 
  $A_1$  &  1 &  0, 4, 6, 8...  \\
  $A_2$  &  1 &  3, 6, 7, 9... \\
  $E$    &  2 &  2, 4, 5, 6... \\
  $T_1$  &  3 &  1, 3, 4, 5, 5...  \\
  $T_2$  &  3 &  2, 3, 4, 5...  \\ 
\hline
\vr
\vr
\vr
\end{tabular}
 \end{table}

\begin{figure}[ht]
\vr
\epsfysize=1.5in
\epsfbox{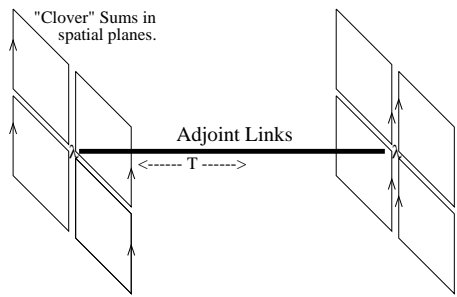}
\vr

\caption{Schematic gluelump sum}
\vr
\vr
\label{fig:dumb}
 \end{figure}

We chose  for simplicity the product of links in a square as the
fundamental object in constructing operators.  This shape does not 
allow all $J^{PC}$ combinations due to cancellations arising from  the
symmetries of the square. Combining the product of links to the adjoint
propagator by  a group generator, diagrammatically the correlation in a
typical $O$ representation looks like that in figure~\ref{fig:dumb}.

\section*{Computational Details}

Measurement of objects containing static propagators are hampered by
cumulative statistical errors from multiplying time links. We observe
that adjoint links are even more sensitive to this effect. In order to
make effective measurements at larger time to determine ground state 
masses, we employ a multi-hit technique. We run a heatbath algorithm
locally on each time link and average over many samples to produce
observables with significant variance reduction.

We measured the correlations for a given state using four paths at both
source and sink. These were constructed using two fuzzing levels on
clover sums built from two sizes of squares. We then employed a
variational technique the resulting matrix to determine the lowest
eigenvalue (See for example \cite{var}). We used a bootstrap analysis to
determine the mass and statistical variation at a time separation of 2
or 3 lattice units, where by inspection of the error on the signal, a
plateau had been reached.

We measured correlations from all sites on various quenched lattices,
the statistics to date are shown in Table~\ref{tab:lattices}. The
fuzzing levels and specific sizes were tuned according to the lattice
spacing to give the best signal.

\begin{table}[ht]
\vrl
\caption{Lattices used in calculation}
\label{tab:lattices}
\begin{tabular}{ccc}
\hline
$\beta$ $\spak$ &   Size $\spak$ &   Number  \\
\hline

  5.7 $\spak$ &   $8^3 \times 16$$\spak$  &   20  \\
  5.7 $\spak$ &   $12^3 \times 24$$\spak$ &   20 \\
  5.9 $\spak$ &   $12^3 \times 24$$\spak$ &   10 \\
  6.0 $\spak$ &   $16^3 \times 48$$\spak$ &   202  \\
  6.2 $\spak$ &   $24^3 \times 48$$\spak$ &   60  \\
\hline

\vr
\vr
\vr
\end{tabular}
\end{table}

\section*{The Adjoint Static Spectrum and Continuum Limit Extrapolation}

\begin{figure}[ht]

\vr
\epsfysize=1.9in
\centerline{\epsfbox{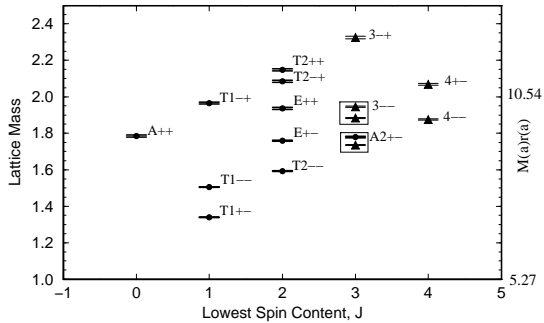}}
\vr
\caption{Mass spectrum of states at $\beta=6.0$.}
\vr
\vr
\label{fig:spec}
\end{figure}

 Figure~\ref{fig:spec} shows the spectrum of states calculated at
$\beta=6.0$. We have used a value $r_0/a(\beta=6.0)=5.272$ to scale the
right hand axis independantly of the lattice spacing $a$. The points
marked by circles are the ten measured representations and are labelled
as such. They are plotted assuming the lowest spin contained in the $O$
representation. 

Note we confirm that the most symmetric state ($A^{++}$) is not the
ground state.

If we consider a variational basis of large dimension, and consider the
restoration of rotational symmetry, then we may expect some of the higher
eigenvalues of a variational analysis to correspond  to higher
spin content in the $O$ representation. We plot some of the excited
states: they are marked them by triangles and labelled according to
their expected (continuum) spin content. Where two states are expected
to coincide in spin content they have been grouped. The spin assignments
shown are in qualitative agreement with this degeneracy. With excited
states added in this way we see clear Regge trajectories beginning to
emerge in the spectrum.

\begin{figure}[ht]
\vr
\epsfysize=1.8in
\centerline{\epsfbox{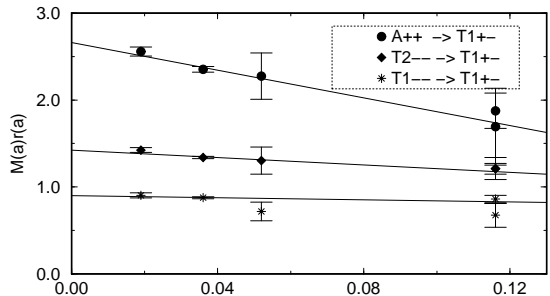}}
\vr
\caption{$\Delta Mr_0$ versus $(a/r_0)^2$, continuum limit extrapolation.}
\vr
\vr
\label{fig:extr}
\end{figure}

A direct extrapolation of the continuum mass spectrum is not feasible as
the self energy of the static propogator has an ultraviolet divergence.
The energy differences between states, however, have a well defined
continuum limit and we extrapolate to $a=0$ the difference between the
second, third and $A^{++}$ states and the lowest $T_1^{+-}$. These
results are shown in table~\ref{tab:cont} and the extrapolation shown in
figure~\ref{fig:extr}. The $r_0$ used to remove explicit $a$ dependence
from observables is also measured in the quenched approximation and
introduces normalization errors $O(10 \%)$.

\begin{table}[tbh]
\vrl
\caption{}
\label{tab:cont}
\begin{tabular}{ccc}
\hline
  Transition &   $\Delta (M_0r_0)$ &   Mev  \\
\hline
  $T_1^{--}$   - $T_1^{+-}$   &   $0.898 \pm 0.022$ &   $354 \pm 9$ \\
  $T_2^{--}$ - $T_1^{+-}$   &   $1.426 \pm 0.023$ &   $562 \pm 9$ \\
  $A^{++}$   - $T_1^{+-}$   &   $2.667 \pm 0.068$ &   $1053 \pm 27$ \\
 \hline
\vr
\vr
\vr
\end{tabular}
\end{table}

As well as $O(a^2)$ errors in measurements that we seek to remove in the
continuum extrapolation, we also note that there may exist  $O(m^2a^2)$
corrections  also. We see in  figure~\ref{fig:spec} that the $T_2^{++}$
and $E^{++}$ states are measured at different energies. We expect, from
their spin content, that in the continuum limit these two states become
degenerate. The extrapolation to $a=0$ of these states is unclear with
our present data, but a common value is within statistical error. 

\section*{Conclusions}

The measured spectrum is in agreement with previous calculations and
phenomenology suggesting that the $1^{+-}$ and $1^{--}$ states of the
gluon field are the lowest lying. The lowest energy difference is higher
than a previous SU(2) calculation which obtained a value of $200 \pm 70$
MeV \cite{lump}.

We note that this is a measurement of the $J^{PC}$ of the gluonic
fields only. The lower lying states correspond to the simplest magnetic
and electric modes of excitation of these fields. With explicit gluino
spin included, these states split into degenerate J=$\half$,
$\threehalfs$ states.

We see that finite size effects are under control and that
extrapolation to the continuum limit of energy differences is well
defined.

The mass spectrum and structure of such states can act as
phenomonological tool to aid understanding of other gluonic states.

\end{document}